# Generating Triplets in Organic Semiconductor Tetracene upon Photoexcitation of Transition Metal Dichalcogenide ReS$_2$

Sourav Maiti[§,*], Deepika Poonia[§], Pieter Schiettecatte,[†,‡] Zeger Hens[†,‡], Pieter Geiregat[†,‡] Sachin Kinge[§,#], Laurens D.A. Siebbeles[§,*]

[§]*Optoelectronic Materials Section, Department of Chemical Engineering, Delft University of Technology, Van der Maasweg 9, Delft, 2629 HZ, The Netherlands*

[†]*Physics and Chemistry of Nanostructures, Ghent University, Ghent, Belgium*

[‡]*Center for Nano and Biophotonics, Ghent University, Ghent, Belgium*

[#]*Toyota Motor Europe, Materials Research & Development, Hoge Wei 33, B-1913, Zaventem, Belgium*

Corresponding Authors: Sourav Maiti and Laurens D.A. Siebbeles

emails: s.maiti@tudelft.nl, l.d.a.siebbeles@tudelft.nl

**Abstract:**

We studied the dynamics of transfer of photoexcited electronic states in a bilayer of the two-dimensional transition metal dichalcogenide $ReS_2$ and tetracene, with the aim to produce triplets in the latter. This material combination was used as the band gap of $ReS_2$ (1.5 eV) is slightly larger than the triplet energy of tetracene (1.25 eV). Using time-resolved optical absorption spectroscopy, transfer of photoexcited states from $ReS_2$ to triplet states in tetracene was found to occur within 5 ps with an efficiency near 38%. This result opens up new possibilities for heterostructure design of two-dimensional materials with suitable organics to produce long-lived triplets. Triplets are of interest as sensitizers in a wide variety of applications including optoelectronics, photovoltaics, photocatalysis, and photon upconversion.

**Table of contents image:**

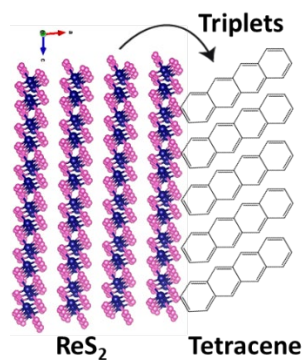

Charge and exciton transfer across heterointerfaces of organic/inorganic semiconductors is both a fundamentally interesting and technologically relevant process for photovoltaics, photocatalysis and optoelectronics.[1-9] Transfer of triplet excitons across such heterointerfaces is of interest as the long triplet lifetimes are advantageous for exciton harvesting and photon-upconversion. Thompson *et al.* and Tabachnyk *et al.* reported transfer of triplet excitons from organic singlet fission molecules to Pb-chalcogenide nanocrystals.[10,11] This result was complemented by Mongin *et al.* who demonstrated reverse triplet transfer from inorganic CdSe nanocrystals to organic surface ligands.[12] To date, triplet transfer has been found to occur in various systems of inorganic nanocrystals and organic molecules, including recently developed perovskite nanocrystals.[13-20] Bulk metal halide perovskites absorbing in the near-infrared have also been utilized as a sensitizer to transfer triplets to organic molecules for triplet-triplet annihilation upconversion.[21,22]

Two-dimensional van der Waals materials, such as transition metal dichalcogenides (TMDCs), are promising for near-infrared light-harvesting followed by transfer of photoexcited electronic states (excitons or free charges) to organic molecules. Two-dimensional TMDC layers can be stacked on top of each other and be mutually bonded through weak van der Waals interactions. Charge transfer across TMDC/organic interfaces has been demonstrated.[1,23-28] Singlet exciton energy transfer from monolayer-$WS_2$ to emissive PbS/CdS nanocrystals has been reported.[29] Kafle *et al.* studied the zinc phthalocyanine (ZnPc)–molybdenum disulfide ($MoS_2$) heterointerface, where photoexcitation of ZnPc results in ultrafast electron transfer to $MoS_2$ creating a charge-separated state. Due to strong spin-orbit coupling in TMDCs, the electron spin flips which results in back electron transfer from $MoS_2$ forming a triplet exciton in ZnPc.[30]

However, to our knowledge the process where photoexcitation of a TMDC, yields triplets in an adjacent organic layer has not been reported yet.

Here we demonstrate, ultrafast (~5 ps) transfer of photoexcited electronic states from the near-infrared absorbing TMDC semiconductor ReS$_2$ (exciton energy of 1.5 eV) to triplet states in tetracene (energy at 1.25 eV) [31,32] in a solid-film bilayer through transient optical absorption spectroscopy. Upon photoexcitation of ReS$_2$, we found distinct signatures of triplets in tetracene that confirm transfer of photoexcited states across the ReS$_2$/tetracene heterointerface. This advocates the applicability of TMDCs, such as ReS$_2$ as a near-infrared triplet sensitizer of organic molecules.

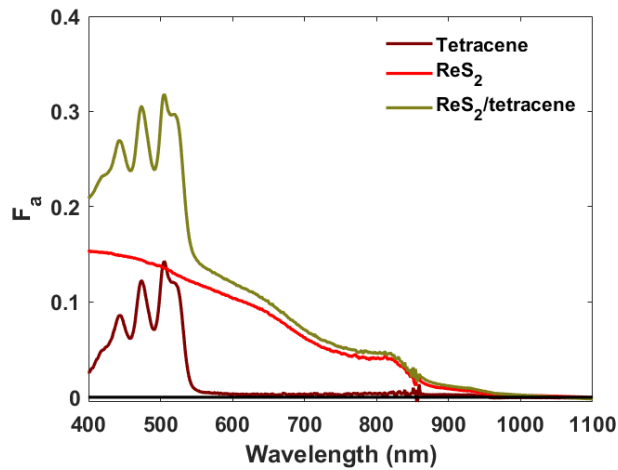

Figure 1. Optical absorption spectra in terms of fraction absorbed photons ($F_a$) of tetracene, ReS$_2$, and ReS$_2$/tetracene films.

Flakes of ReS$_2$ were obtained by liquid-phase exfoliation, as reported in our previous paper.[33] The ReS$_2$ flakes have a thickness of 4±2 monolayers and average lateral sizes of 75 nm (Figure S1, SI). We refer to Figure 1, SI for additional details. A dispersion of ReS$_2$ flakes dispersed in N-methyl-2-pyrrolidone (NMP) was drop-cast on a quartz substrate and the NMP was evaporated under a nitrogen atmosphere inside glovebox forming a film. A ReS$_2$/tetracene bilayer was obtained by deposition of tetracene on top of a ReS$_2$ film through thermal evaporation. Figure 1 shows the optical absorption spectra of tetracene, ReS$_2$, and the ReS$_2$/tetracene bilayer. Pure tetracene has a structured absorption spectrum in which the lowest energy singlet-exciton split at 505 nm and 520 nm (Davydov splitting).[34,35] ReS$_2$ has a first excitonic absorption near 825 nm (Figure 1). The bilayer clearly shows the features associated with the individual components.

To monitor the dynamics of exciton or charge transfer from ReS$_2$ to tetracene broadband (500-900 nm) femtosecond transient absorption (TA) spectroscopy was utilized. The experimental setup has been described elsewhere.[36] Here we monitor the pump induced changes in the absorption of the probe pulse ($\Delta A = A_{pump\ on} - A_{pump\ off}$, where $A$ is the absorbance) as a function of pump-probe delay time at different probe photon energies. The TA signal can exhibit negative $\Delta A$ signals due to depletion of the ground state (ground state bleach, GSB) and/or stimulated emission (SE) from excitons. Also, photoinduced absorption (PIA) or positive $\Delta A$ in TA can arise when pump-generated charge carriers and/or excitons get excited further by absorbing probe photons. The $\Delta A$ signal can be expressed as

$$\frac{|\Delta A|}{I_0 F_a} = \varphi \frac{\sigma_B}{\ln 10} \qquad , \qquad (1)$$

where $I_0$ is the number of incident photons per unit area, $F_a$ is the fraction absorbed pump photons in the sample, $\varphi$ is the quantum yield of charges/excitons and $\sigma_B$ is the cross-section of bleach or photoinduced absorption.

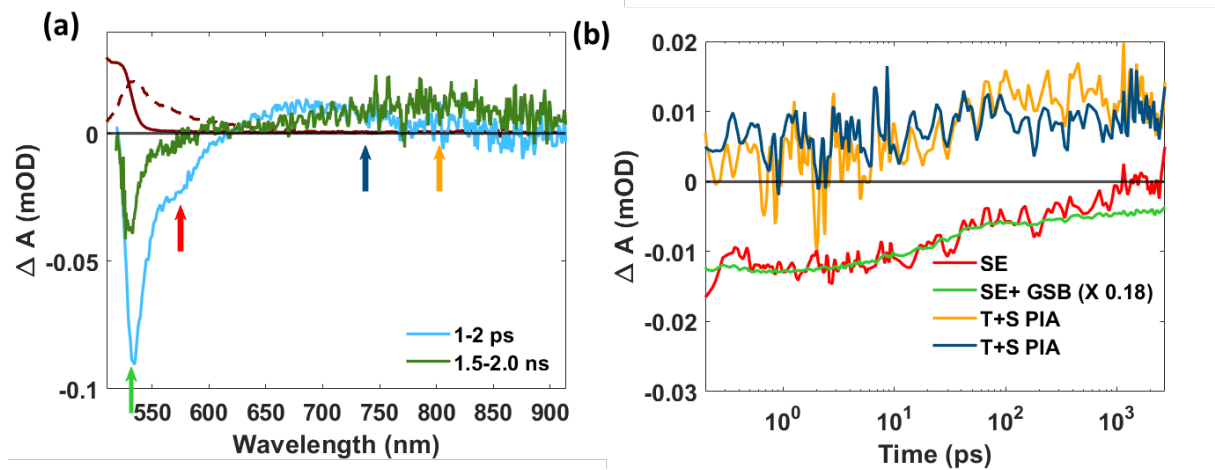

Figure 2. (a) TA spectra of tetracene at early (1-2 ps) and longer (1.5-2.0 ns) time delays after the 510 nm pump laser pulse. The steady-state absorption (dark red) and photoluminescence (dotted dark red) spectra are also shown. The colored arrows indicate the wavelength region where decay kinetics were monitored. (b) Normalized decay kinetics (on a log scale) for four different wavelength regions; stimulated emission (SE) and ground state bleach (GSB) at 530-545 nm, stimulated emission (SE) at 580-600 nm from singlets (S), photoinduced absorption (PIA) at 720-765 nm and 780-820 by both singlets (S) and triplets (T). The TA kinetics has been averaged over the wavelength regions.

Below we first discuss the origin of the pump-induced changes in the absorption of the probe pulse for pure tetracene. We identify wavelength regions in terms of GSB/SE and PIA corresponding to singlets and triplets. That knowledge is used to interpret the TA results ReS$_2$/tetracene bilayer.

**Tetracene.** Figure 2(a) shows the evolution of the TA spectrum of tetracene upon 510 nm photoexcitation with an absorbed photon density of $2.1 \times 10^{12}$ cm$^{-2}$ at which exciton-exciton annihilation is insignificant.[37] The salient features at early delay times (1-2 ps) in Figure 2(a) are negative $\Delta A$ (bleach) around 530-545 nm (green arrow) and 580-600 nm (red arrow). These wavelengths correspond to the 0-0 and 0-1 photoluminescence peaks of tetracene[34] and we, therefore, attribute these bleach features in the TA spectrum to stimulated emission from singlets. At 530-545 nm ground-state depletion will also contribute to bleach. As shown in Figure 2(b), both bleach features decay on a time scale of 100 ps due to singlet fission, in agreement with previous reports.[34, 38] The bleach at 580-600 nm has completely decayed after 1000 ps, which we attribute to the disappearance of SE. By contrast, the bleach at 530-545 nm persists, due to ground state depletion by triplets with a lifetime as long as ~100 ns.[39,40]

The TA spectra exhibit broad PIA at wavelengths longer than 650 nm. The PIA at 650-665 nm due to singlets decays on a timescale of 100 ps (Figure S2, SI).[34] The PIA at 780-820 nm is due to both singlets and triplets being excited to higher-lying states of the same spin. The ratio of the oscillator strength of the PIA for singlets and triplet pairs is close to 1:2.[34] The increase of the PIA at 780-820 nm in Figure 2(b) reflects the dynamics of triplet formation via singlet fission (see also Figure S2, SI). In the region 720-765 nm, the PIA is virtually constant during measurement time, which we attribute to the PIA by a singlet exciton being similar to that by a triplet pair.

**ReS₂/tetracene bilayer**. We studied transfer photoexcited electronic states from ReS$_2$ to tetracene in a bilayer configuration. Please note, according to our recent study photoexcitation at 530 nm can lead to the formation of excitons and free charges in ReS$_2$ with an upper limit of the initial quantum yield of free charges near 20% and a lifetime before recombination to excitons or trapping of a few picoseconds.[41] After photoexcitation of ReS$_2$ triplets can be produced in tetracene both due to transfer of excitons or sequential transfer of electrons and holes from ReS$_2$ to tetracene. In this work, we do not distinguish these two processes and henceforth refer to them as transfer of photoexcited states to tetracene. The ReS$_2$/tetracene bilayer was photoexcited at 510 nm (above the band gap of tetracene) and at 700 nm (below the band gap of tetracene). The pump laser entered the ReS$_2$/tetracene bilayer from the side of tetracene. We first discuss the data obtained at 510 nm from which we infer transfer of photoexcited states from ReS$_2$ to triplet states in tetracene. The 700 nm photoexcitation data corroborate this mechanism of triplet generation in tetracene after photoexcitation of ReS$_2$.

Since pump laser light of 510 nm is absorbed both by tetracene and ReS$_2$, it leads to the initial formation of both singlets in tetracene and photoexcited states in ReS$_2$. Using the data in Figure 1, we calculated the number of photons absorbed per unit area in tetracene ($I_0 F_a^{tet}$) and ReS$_2$ ($I_0 F_a^{ReS_2}$) separately and determined the corresponding $\Delta A / I_0 F_a$, which is directly proportional to the quantum yield, $\varphi$, of excitons probed at a particular wavelength, see Equation 1. Details of the calculations can be found in section S7, SI. Figure 3(a) shows that the SE bleach (580-600 nm) due to singlets in tetracene is identical in both tetracene and the bilayer. This indicates that initially hot photoexcited states at 510 nm in ReS$_2$ (well-above the exciton energy of 1.5 eV) relax quickly to lower energies in ReS$_2$, preventing the significant transfer of hot electron-hole pairs to singlet states in tetracene. It can be seen in Figure S3, SI that the time-resolved PIA at 495-500 nm due to singlet excitons (we can ascribe the fast decay to singlets in tetracene)[42] in tetracene is similar for tetracene and the ReS$_2$/tetracene bilayer. This corroborates the absence of production of singlets in tetracene from hot photoexcited states in ReS$_2$.

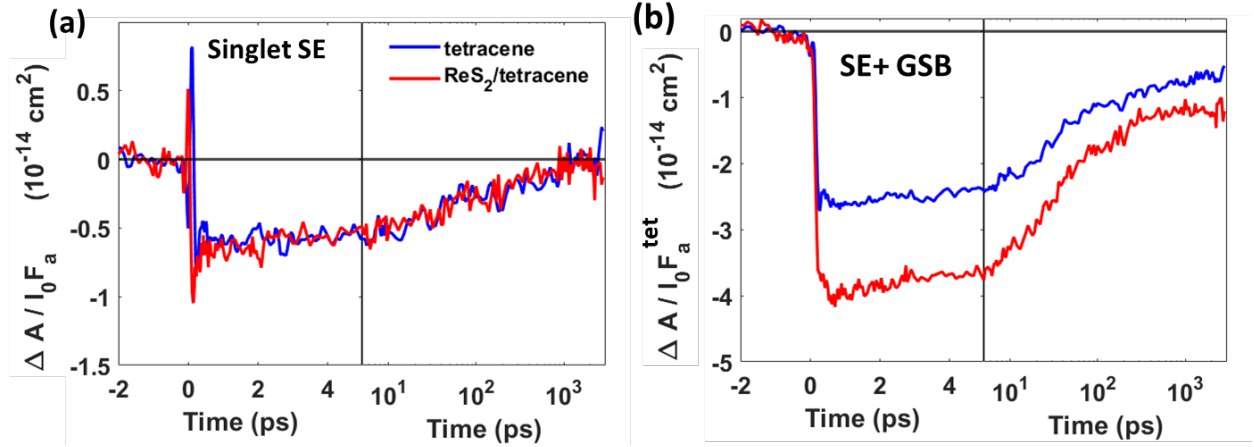

Figure 3. $\Delta A/I_0 F_a^{tet}$ after 510 nm photoexcitation of the tetracene film (blue) and the ReS$_2$/tetracene bilayer (red), due to (a) SE from singlets in tetracene (580-600 nm) and (b) SE and GSB (530-545 nm) due to singlets and triplets in tetracene.

Interestingly, Figure 3b shows that $\Delta A/I_0 F_a^{tet}$ at 530-545 nm due to both SE and GSB from singlets and triplets in tetracene is higher for the ReS$_2$/tetracene bilayer than for tetracene only. We attribute this difference to the transfer of photoexcited states from ReS$_2$ to triplets in tetracene. Note, that pure ReS$_2$ does not have any significant signal in this wavelength region, as shown in Figure S4, SI. Using the magnitudes of $\Delta A/I_0 F_a^{tet}$ for tetracene and the bilayer, the efficiency of photoexcited state transfer from ReS$_2$ to tetracene was calculated to be 38±13%, assuming a triplet yield of 1.5[43], for details see section S7, SI. We also monitored the time-dependent population of triplets in tetracene via the PIA at 720-765 nm, see Figure S3, SI. The magnitude of $\Delta A/I_0 F_a^{tet}$ is higher (with a rise till 5 ps) for the bilayer than for tetracene, which further substantiates photoexcited state transfer from ReS$_2$ to triplets in tetracene. The absence of decay of the PIA on the 2 ns timescale in Figure S3, SI agrees with the much longer triplet lifetime of ~100 ns.[39,40]

The TA kinetics at the ReS$_2$ ground state bleach position (830-840 nm) further corroborates the transfer of photoexcited states from ReS$_2$ to triplets in tetracene. The TA spectrum of pure ReS$_2$ (Figure S5, SI) upon 510 nm photoexcitation exhibits bleach (negative $\Delta A$) around 835 nm at early time delays, due to ground state bleach. The bleach signal recovers fast and changes into PIA due to charge carriers in ReS$_2$.[44] As shown in Figure 4a the $\Delta A/I_0 F_a^{ReS_2}$ at a short time (< 1 ps) (see Figure S5, SI), due to photoexcited states in ReS$_2$ is similar for the ReS$_2$ film and the ReS$_2$/tetracene bilayer. Interestingly, the PIA on longer times due to photoexcited states in ReS$_2$ (see Figure 4a) is lowest for the bilayer (by 57±13%). This is an additional indication of transfer of photoexcited states from ReS$_2$ to long-lived triplets in tetracene.

Now we consider selective photoexcitation of ReS$_2$ at 700 nm. Figure 4b shows that PIA at 830-840 nm due to photoexcited states in ReS$_2$ is reduced in the bilayer by 51±15%, which we attribute to transfer of photoexcited states from ReS$_2$ to tetracene, in agreement with the results for 510 nm photoexcitation in Figure 4a. Moreover, Figure S6 shows that the PIA at 720-765 nm due to triplets in tetracene increases during the first 5 ps and persists longer than 2 ns, in agreement with the much longer triplet lifetime of ~100 ns.[39,40]

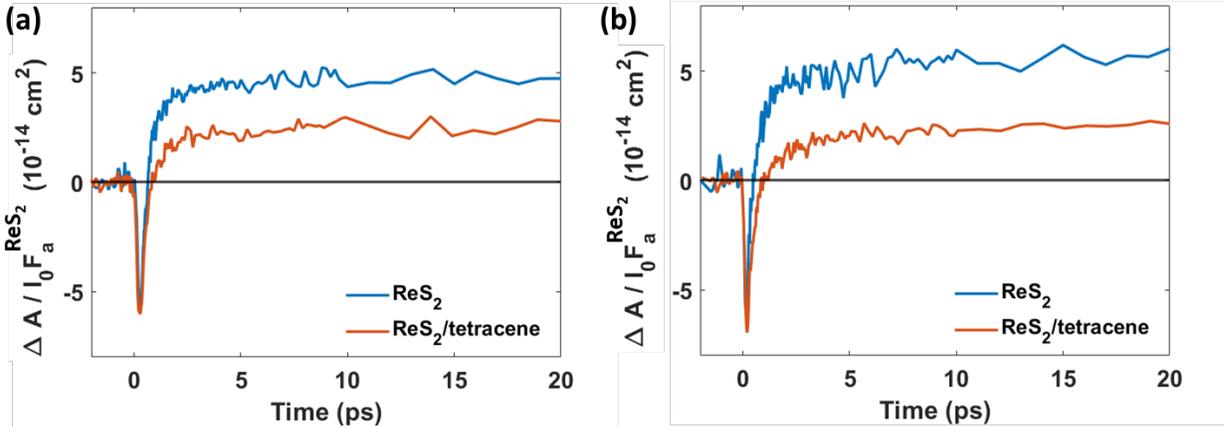

Figure 4. $\Delta A/I_0 F_a^{ReS_2}$ at the ReS$_2$ ground state bleach position (830-840 nm) for ReS$_2$ and ReS$_2$/tetracene after (a) 510 nm and (b) 700 nm photoexcitation.

In conclusion, we have shown transfer of photoexcited states from a solution-processed ReS$_2$ layer to tetracene on a timescale of 5 ps. This is orders of magnitude faster than triplet transfer from semiconductor nanocrystals or perovskite to organic molecules. This opens up the possibility of triplet-triplet annihilation upconversion in organic materials upon light absorption in inorganic TMDCs. Future research is needed to increase the efficiency of the transfer of photoexcited states by improving the electronic coupling between the TMDC layer and the organic material.

**Acknowledgments**

We thank Dr. Kevin Felter for his support during tetracene deposition by thermal evaporation. This research received funding from the Netherlands Organisation for Scientific Research (NWO) in the framework of the Materials for sustainability and from the Ministry of Economic Affairs in the framework of the PPP allowance. P.S. thanks the FWO Vlaanderen for funding.

**Supporting Information paragraph**

Characterization of ReS$_2$, additional TA data, and transfer efficiency calculations.

# Supporting Information

# Generating Triplets in Organic Semiconductor Tetracene upon Photoexcitation of Transition Metal Dichalcogenide ReS$_2$

Sourav Maiti[§], Deepika Poonia[§], Pieter Schiettecatte,[†,‡] Zeger Hens[†,‡], Pieter Geiregat[†,‡] Sachin Kinge[§,#], Laurens D.A. Siebbeles[§]

[§]*Optoelectronic Materials Section, Department of Chemical Engineering, Delft University of Technology, Van der Maasweg 9, Delft, 2629 HZ, The Netherlands*

[†]*Physics and Chemistry of Nanostructures, Ghent University, Ghent, Belgium*

[‡]*Center for Nano and Biophotonics, Ghent University, Ghent, Belgium*

[#]*Toyota Motor Europe, Materials Research & Development, Hoge Wei 33, B-1913, Zaventem, Belgium*

Corresponding Authors: Sourav Maiti and Laurens D.A. Siebbeles

emails: s.maiti@tudelft.nl, l.d.a.siebbeles@tudelft.nl

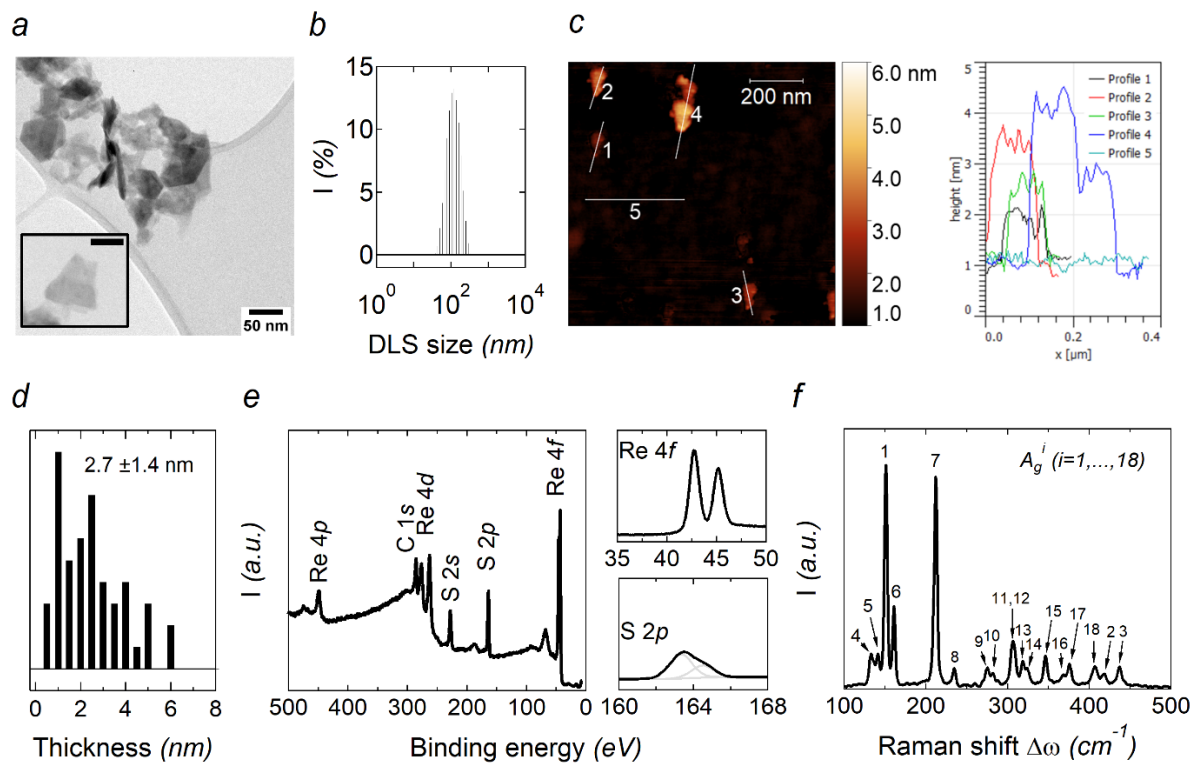

Figure S1. Overview of the material characteristics of liquid-phase exfoliated rhenium disulfide. (a) TEM image. (b) DLS size distribution. (c) AFM images together with extracted line profiles. (d) AFM histogram. (e) XPS survey spectrum together with high-resolution XPS spectra of the Re 4f and S 2p core levels. (f) Raman spectrum with the $A_g^i$ modes of $ReS_2$ labeled according to the literature.[1]

Characterization of liquid-phase exfoliated rhenium disulfide. We exfoliated $ReS_2$ in *N*-methyl-2-pyrrolidone (NMP) following a procedure detailed in our previous work.[2] Transmission electron microscopy (TEM, Figure S1a) reveals thin $ReS_2$ flakes with lateral dimensions of a few tens to a hundred nanometer. Dynamic light scattering (DLS, Figure S1b) corroborates this result by showing a monomodal particle size distribution centered around 106 nm, a value corresponding to an estimated lateral size of ≈ 75 nm according to the DLS sizing curves for two-dimensional materials proposed by the Coleman group.[3] Thickness measurements by atomic force microscopy (AFM, Figure S1c) indicate that the flakes are typically a few nanometers thick, corresponding to a layer number of ≈ 4±2 ReS$_2$ layers (see AFM histogram, Figure S1d). X-ray photoelectron spectroscopy (XPS, Figure S1f) emphasizes that the flakes are stoichiometric and oxide-free. This result is supported by Raman spectroscopy (Figure S1g) that only shows the 18 characteristic vibrations of ReS$_2$[1]. We refer the reader to our previous work for a more elaborate description.[2]

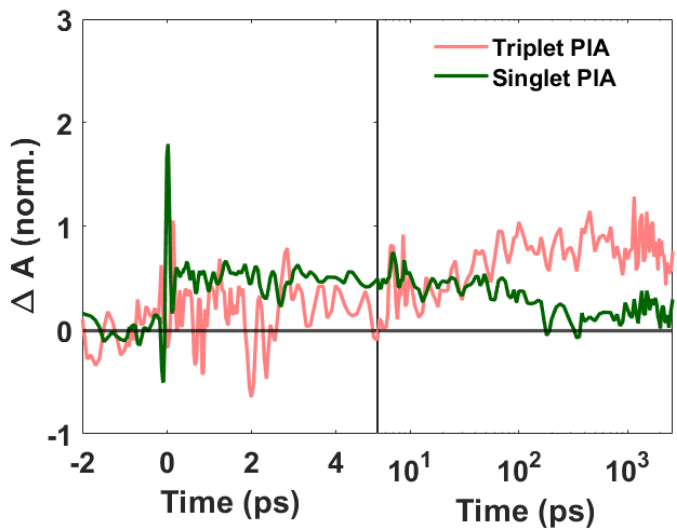

Figure S2. Tetracene photoinduced absorption (PIA) kinetics (normalized to maxima at a short time) in the range 650-665 nm (singlet PIA) and 780-820 nm (triplet PIA, also shown in Figure 2(b) of main text).

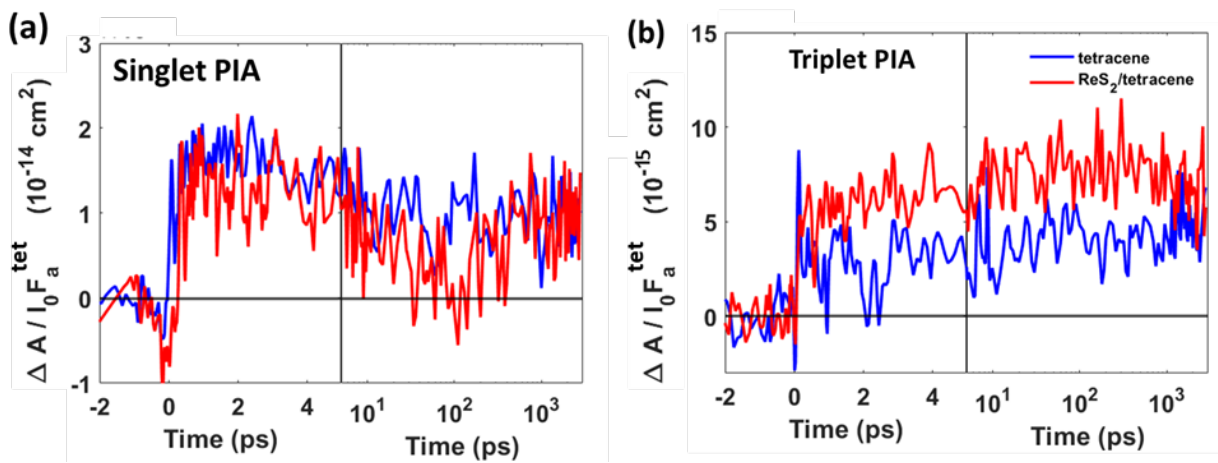

Figure S3. Comparison of time-dependent TA for the tetracene film and the ReS$_2$/tetracene bilayer due to (a) absorption by singlets in tetracene (495-500 nm) and (b) triplets in tetracene (720-765 nm).

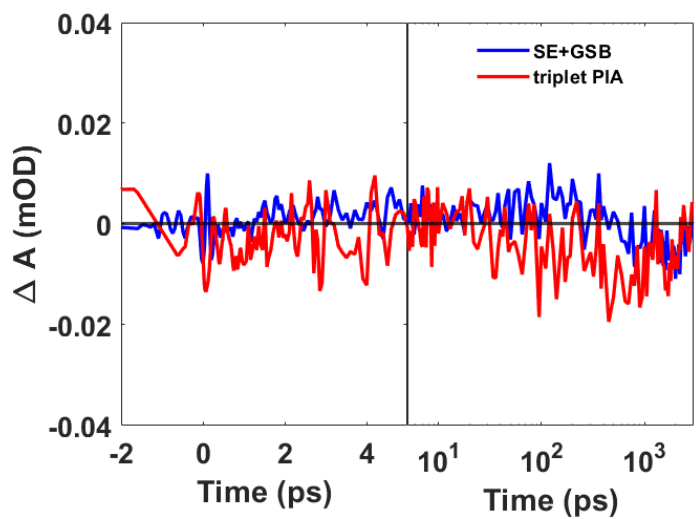

Figure S4. ΔA for the ReS$_2$ film at wavelengths where tetracene exhibits ground state bleach and stimulated emission (530-545 nm) or triplet photoinduced absorption (PIA at 720-765 nm) after 510 nm photoexcitation.

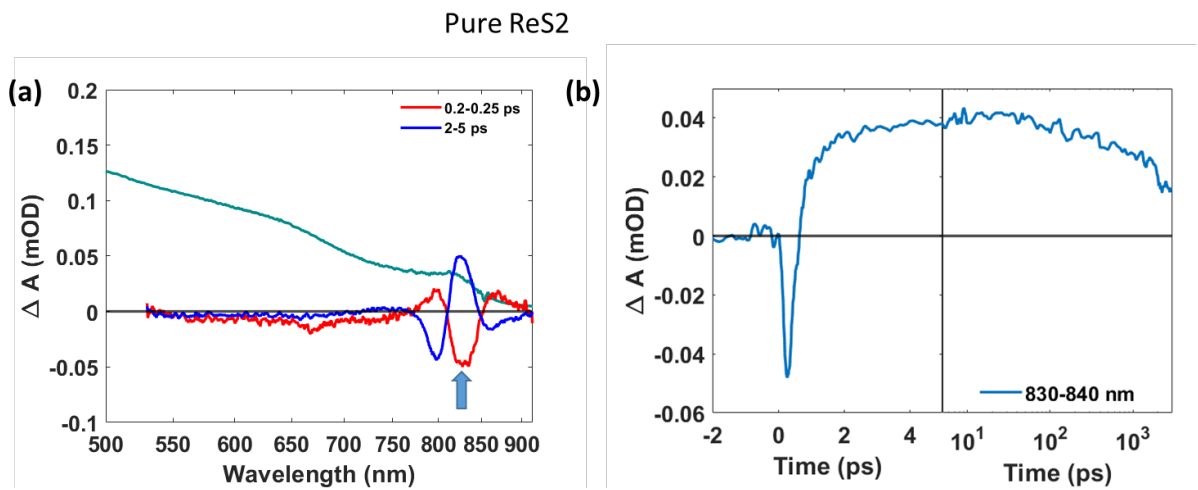

Figure S5. (a) TA spectra (after 510 nm photoexcitation) of the ReS$_2$ film at 0.20-0.25 ps and 2-5 ps delay time, together with the ground state absorption in dark cyan. (b) Kinetics in the probe region 830-840 nm. At longer delay times (> 2 ps) the TA spectra have a pronounced PIA near 820 nm along with bleach features at shorter and longer wavelengths, which has been attributed to pump-induced shifts and broadening of the optical absorption spectrum.[4]

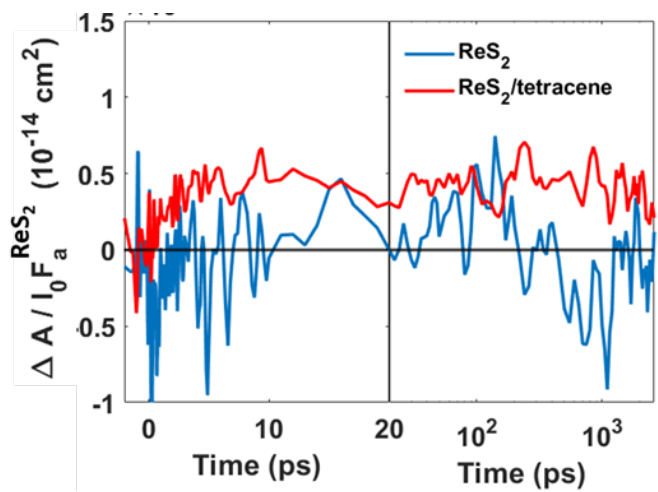

Figure S6. $\Delta A/I_0 F_a^{ReS_2}$ after 700 nm photoexcitation at tetracene triplet photoinduced absorption (720-765 nm) of the ReS$_2$ film (blue) and the ReS$_2$/tetracene bilayer (red).

**S7: Calculation of exciton transfer efficiency:**

**510 nm pump excitation**

For the tetracene/ReS$_2$ bilayer: photoexcitation through tetracene side

$$I_0 = 1.7 \times 10^{13} cm^{-2}$$

In the bilayer,

$$F_a^{tet} = F_a^{total} - F_a^{ReS_2} = 0.31 - 0.14 = 0.17$$

$$I_0 F_a^{tet} = (1.7 \times 10^{13}) \times 0.17 = 2.9 \times 10^{12} cm^{-2}$$

$$I_0 F_a^{ReS_2} = (I_0 - I_0 F_a^{tet}) \times F_a^{ReS_2} = (1.4 \times 10^{13}) \times 0.14 = 2 \times 10^{12} cm^{-2}$$

**At triplet GSB positon (530-545 nm) position**

For pure tetracene, the GSB between 530-545 nm from 500 ps-3 ns (when all the singlets are converted to triplets)

Using equation 1: $\frac{|\Delta A|}{I_0 F_a} = \varphi \frac{\sigma_B}{\ln 10}$

$\varphi = 1.5$ for singlet fission.

$\sigma_{GSB}$ is calculated considering the pure tetracene TA data for 510 nm excitation.

$$\sigma_{GSB} = \frac{|\Delta A| \times \ln 10}{\varphi \times I_0 F_a^{tet,pure}} ; (\Delta A = 26.9 \pm 2.8 \, \mu OD)$$

In the bilayer,

$$\Delta A^{Bi} = \Delta A^{tet} + \Delta A^{tr} = \frac{\varphi \sigma_{GSB} I_0 F_a^{tet} + \varphi_{tr} \sigma_{GSB} I_0 F_a^{ReS_2}}{ln10}$$

$$\Delta A^{Bi} = 58.5 \pm 4.2 \; \mu OD$$

$\varphi_{tr} = 0.38 \pm .13$